\documentstyle[a4,12pt]{article}

\begin{document}

\begin{titlepage}
\rightline {Si-95-2\  \  \  \   }

\vskip 3.truecm

\centerline{\large \bf The short-time Dynamics of the
  Critical Potts Model}
\vskip 2.0truecm
\centerline{\bf  L. Sch\"ulke and B. Zheng }

\vskip 0.7truecm

\centerline{Universit\"at-GH Siegen, D-57068 Siegen, Germany}

\vskip 1.truecm

\centerline{\large March 1995}

\vskip 2.7truecm

\abstract{
	The universal behaviour of the short-time dynamics
of the three state Potts model in two dimensions at
criticality is investigated with Monte Carlo methods.
 The initial increase of
the order is observed. The new dynamic exponent $\theta$
as well as exponent $z$ and
$\beta / \nu$  are determined.
The measurements are carried out in the very beginning
of the time evolution. The spatial correlation length
is found to be very short compared with
the lattice size. }


\end{titlepage}

for years it has been known that there exist universality and scaling
behaviour for statistical systems at criticality in equilibrium or near
equilibrium, more or less due to the {\it infinite }
spatial and time correlation length.
Recently it has been observed
that universality and scaling may also be present
far from equilibrium. One of the examples is that the Ising model,
initially in very high temperature, is suddenly quenched
to the critical temperature and then evolves with the dynamics of model A.
According to an argument of Janssen et al. with
two-loop $\epsilon$-expansion \cite {jan89},
besides the well-known long-time universal behaviour, there exists another
universal stage in the earlier time, termed {\it critical
initial slip}, which sets in right after the microscopic time
scale. The characteristic behaviour of such process is that,
if a non-zero but {\it small} initial magnetization $m_0$
is generated in the system,
the anomalous dimension of the operator $m_0$ gives rise to the
critical increase of the magnetization
\begin{equation}
M(t) \sim m_0 \, t^\theta,
\label{cis}
\end{equation}
with $\theta$ being a new dynamic critical exponent.
Detailed scaling analysis reveals \cite{die93}
that the characteristic time scale
for the critical initial slip is
$t_0 \sim m_0 ^ {-z/x_0}$,
where $x_0$ is the scaling dimension of $m_0$, and related to $\theta$
by
$x_0 =  \theta z + \beta \nu$.
Interesting enough, it was pointed out that the exponents $\beta$,
$\nu$ and $z$ should be valued  the same as those in the
equilibrium or long-time stage of the relaxation.
Previously $\theta$  has been measured
with Monte Carlo simulation in two dimensions somehow {\it indirectly}
from the power law decay of the autocorrelation
\cite{hus89,bra91}, and recently in three diemensions
{\it directly} from the power law
increase of the magnetization \cite{li94}.
They are in good agreement with the result
from the $\epsilon$-expansion.
Furthermore, based on the scaling relation in the initial stage
of the time evolution, a new promising way for measuring the
exponents $z$, $\beta$ and $\nu$
from the finite size scaling
has been proposed \cite{li94a}.
This indicates a possible broad application of the short-time
dynamics. Therefore more and deeper understanding
of this phenomenon becomes urgent.

As far as we know, even though analytical
perturbative calculations
for the critical initial slip can be extended to the
$O(N)$ vector model \cite{jan89,die93},
the numerical results are so far limited
to the Ising model \cite{hus89,bra91,men94,li94,li94a}.
The purpose of this letter is to report
 systematic Monte Carlo
simulations of the short-time behaviour
of the two dimensional Potts model at criticality
relaxing from high temperature states.
 A refined measurement of the
exponent $\theta$, $z$ and $\beta / \nu$
from the power law behaviour of the physical observables
is presented.
It relates the above mentioned
indirect and direct measurements
of $\theta$ each other and provides a consistent
test of the scaling relation for the Potts model.
Finite size effects and the week dependence of the
measurement of  $\theta$ on $m_0$ are discussed and the
spatial correlation length is computed.

{}From the scaling analysis,
the autocorrelation has the initial behaviour
\begin{equation}
A(t) \sim t^{-d/z+\theta}
\label{auto}
\end{equation}
in case of $m_0 = 0$. Most of the previous measurements
of $\theta$ were based on this relation.
A disadvantage is that one has to input $z$ to
obtain $\theta$. Since $z$ is one order of magnitude bigger
than $\theta$, a small relative error of $z$ will induce
a big error in $\theta$. This becomes more severe when $\theta$
is getting smaller. A direct measurement from the power law
increase of the order parameter can improve
this situation \cite{li94}. As we will see later,
 for the Potts model we did observe the same initial
increase of the order shown in (\ref{cis}).
The week dependence of the practical
measurement of  $\theta$  on $m_0$ appears, however,
more visible in the Potts model
than in the Ising model. Therefore in this letter we consider a
correction of $\theta$ for finite $m_0$.
Since  $\theta$ for the Potts model is smaller than that
for the 2D Ising model, its measurement from the auto-correlation
is harder.

On the other hand, traditionally the
dynamic exponent $z$ is defined and
 measured from the long-time exponential decay
of the time correlation or the magnetization
of the systems \cite{wil85,wan91}. Due to the critical slowing
down this is somehow difficult. From the above discussion,
however, it is easy to
realize that with $\theta$ in hand, one can obtain $z$ quite
rigorously from the power law decay (\ref{auto})
of the autocorrelation.
Since the measurement is carried out
at the beginning of the time evolution, it is efficient.
This is an alternative way to measure $z$ from the short-time
behaviour of the system. Compared with the method proposed
in \cite{li94a}, the advantage is that the measurement can be
carried out in a single lattice rather than by comparing two
lattices. Finally the static exponent $\beta / \nu$ can be obtained
from the power law increase of the second moment
\begin{equation}
M^{(2)}(t) \sim t ^ {(d-2\beta / \nu)/z}.
\label{m2}
\end{equation}
Since $2\beta / \nu$ is one order of magnitude
smaller than $z$, its measurement is quite
sensitive to the error of $z$.

\begin{figure}[t]\centering
\caption{
Magnetization $M(t)/M(0)$ vs. time in double-log scale
for the initial value $m_0 = 0.08$.
}
\label{F1}
\end{figure}

\begin{figure}[h]\centering
\caption{
Autocorrelation $A(t)$ vs. time in double-log scale  for $L=9$, $18$, 3$6$,
$72$, $144$ and $288$.
}
\label{F2}
\end{figure}

For the study of the short-time dynamics,
 usually quite a big lattice is used.
In this letter, the finite size effects will be
discussed. It turns out that for the measurement
of $\theta$ the finite size effect is not so big.
Furthermore, the spatial
correlation is measured and found to be very small
compared with the lattice size. This indicates that
the mechanism for the universality and scaling
in short-time dynamics should be different from that in the
equilibrium or near equilibrium.

The Hamiltonian for the $q$ state Potts model
\begin{equation}
H=J  \sum_{<ij>}  \delta_{\sigma_i,\sigma_j},\qquad \sigma_i=1,...,q
\label{hami}
\end{equation}
with $<ij>$ representing nearest neighbors.
In this letter we only consider the three state case.
It is well known that for the three state Potts model the
 critical point locates at $J_c=\log(1+\surd 3)$.
As in case of the Ising model,
initially the Potts model
is prepared to be in a random state
with a sharp magnetization $m_0$. Then it is released
to the evolution with the heat-bath algorithm at the
critical temperature. We measure the time evolution
of the magnetization, the second moment, the auto-correlation
and the spatial correlation respectively
\begin{equation}
M(t)= \frac{3}{2}\frac{1}{N}\,
   \left<\sum_i \left(\delta_{\sigma_i(t),1}-\frac{1}{3}\right)\right>,
\end{equation}
\begin{equation}
M^{(2)}(t)= \frac{9}{4}\frac{1}{N^2}\,
   \left<\left[\sum_i
\left(\delta_{\sigma_i(t),1}-\frac{1}{3}\right)\right]^2\right>,
\end{equation}
\begin{equation}
A(t)=\frac{1}{N}\,
   \left<\sum_i
\left(\delta_{\sigma_i(0),\sigma_i(t)}-\frac{1}{3}\right)\right>,
\end{equation}
\begin{equation}
C(x,t)=\frac{1}{N}\,
   \left<\sum_i
\left(\delta_{\sigma_i(t),\sigma_{i+x}(t)}-\frac{1}{3}\right)\right>,
\end{equation}
where the average is taken over the independent
intial configurations.
Except for the magentization $M(t)$, the above
definitions are restricted here to the case of $m_0=0$.
In spite of the lack of the
analytic analysis, we assume that all the scaling properties
including the increase of the order
for the Ising model are valid also for the Potts model
and test them by numerical simulation.
In the same time the related critical exponents will
be determined.

\begin{figure}[t]\centering
\caption{
Second moment $M^{(2)}(t)/M^{(2)}(1)$ vs. time in double-log scale.
}
\label{F3}
\end{figure}

In Fig.~1, as an example,
 the time evolution of the magnetization with
the initial value $m_0=0.08$ for different lattice size $L$
is displayed in a double log-scale to present
the power law increase.
It is remarkable that the power law increase starts also
from the very beginning
of the time evolution $t=1$ as it is in the three dimensional
Ising model.
$\theta$ can be estimated from
the slope of the curves. It is clearly seen that
$\theta$ converges to a definite value when the
lattice size $L\geq 36$.
In other words, for the measurement of $\theta$ the finite
size effect is already quite small for a lattice size $L=36$.
In comparison to this, to observe the power law decay
for the auto-correlation one needs much bigger lattices,
as will be seen later.
In Tab.~1, $\theta$ for $L=72$ measured from
$t=1$ to $t=15$ for different
initial magnetizations has been summarized.
The total number of samples for the independent initial
configurations is $80,000$ for bigger $m_0$ and
$480,000$ for smaller $m_0$. The errors are estimated
by dividing the data into four or six groups, respectively.
Different from the Ising model,
the measured $\theta$ shows slightly more dependence on $m_0$.
Therefore,  according to its definition, a linear extrapolation of
$\theta$ to the fixed point $m_0=0$ is carried out.
This leads to
the value
$$
\theta=0.0815(27).
$$

\begin{table}[t]\centering
$$
\begin{array}{|c|l|l|l|l|l|}
\hline
 m_0  &\qquad 0.10 &\quad 0.08 &\quad 0.06 &\quad 0.04 &\quad 0.00\\
\hline
\theta & 0.1076(08) & 0.1036(12) & 0.0980(06) & 0.0925(14) & 0.0815(27) \\
\hline
\end{array}
$$
\caption{ The measured $\theta$ for $L=72$.}
\label{T1}
\end{table}

\begin{table}[h]\centering
$$
\begin{array}{|c|r|r|r|r|}
\hline
 L  &\qquad 72 &\quad 144 &\quad 288 &\quad \infty\\
\hline
-d/z+\theta & -0.8510(08) &  -0.8387(09)& -0.8335(09) &  -0.8283(20)\\
(d-2\beta/\nu)/z & 0.7921(11) & 0.7881(14) & 0.7875(28) & 0.7878(16)\\
\hline
\end{array}
$$
\caption{ Results for $L=72, 144$ and $288$.}
\label{T2}
\end{table}

Now we set $m_0=0$ and measure the auto-correlation.
In Fig.~2, the dependence of the auto-correlation on $L$
is presented. Obviously at $L=36$ no power law behaviour is
observed. The convergence to a power law behaviour
only starts around $L=144$.
It is clear that the regime presenting power law grows
when the lattice size increases.
It is interesting that,
somehow different from $M(t)$, the first time steps
apparently deviate slightly from the power law.
In Tab.~2 the corresponding values for $-d/z+\theta$
measured from $t=5$ to $t=50$ are given.
We stop the measurement at $t=50$ due to the obvious bigger
finite size effect and statistical errors.
The total samples for
$L=144$ is $40,000$ and for $L=288$ is $16,000$.
if we only intend to obtain $z$ by taking
$\theta$ as input, we should already be satisfied with the lattice size
since the results from $L=144$ and $L=288$ are already so close.
However, in order to get the better $\beta/\nu$ later
by inputting the $z$ measued here,
 we perform the
linear extrapolation for $-d/z+\theta$ over $1/L$ to
$L=\infty$ and obtain
$$
z=2.1983(81).
$$
Compared to the values of $z$ distributed between $z=2.2$ and $z=2.7$
from different numerical measurments \cite{a,TANG87,b,c}, our result supports
the relative small $z$ \cite{a,TANG87}.

In Fig.~3,  the power law increase of the second moment
$M^{(2)}(t)$ is shown. The measurement of
 $(d-2\beta/\nu)/z$ from $t=5$ to $t=50$ are also
listed in Tab.~2.
{}From the data we can see that for $L=144$ and $L=288$,
the finite size effects are already less prominent than
the statistical errors.
Therefore the value $0.7878(22)$ of $(d-2\beta/\nu)/z$ at
$L=\infty$ is only a simple average of them.
Here we get
 $$
 2\beta/\nu=0.2682(73),
 $$
which is in good agreement with the exact value
$4/15\approx 0.2667$ \cite{bax82}.
Such coincidence provides a strong
support for scaling in the short-time dynamics.

Finally we have also measured the correlation length $\xi(t)$
from the spatial correlation function $C(x,t)$.
for example,  for $L=288$,
$\xi(t=96) \approx 6.0$ and it is much smaller than the lattice size $L$.

In conclusion, by observing the power law behaviour
of the magnetization $M(t)$, the second moment $M^{(2)}(t)$
and the auto-correlation $A(t)$, we confidently confirm
the scaling properties for the Potts model in the short-time
dynamics and obtain the related critical exponents
$\theta$, $z$ and $\beta/\nu$.
This is the first measurment of $\theta$ for the Potts model.
Our way to determine $z$ is efficient.
Such an investigation
for models in other universality classes should be carried out.

\vskip 1. truecm

{\it Acknowledgement:} The authors would like to thank
   K. Untch for the help in maintiaining our Workstations.

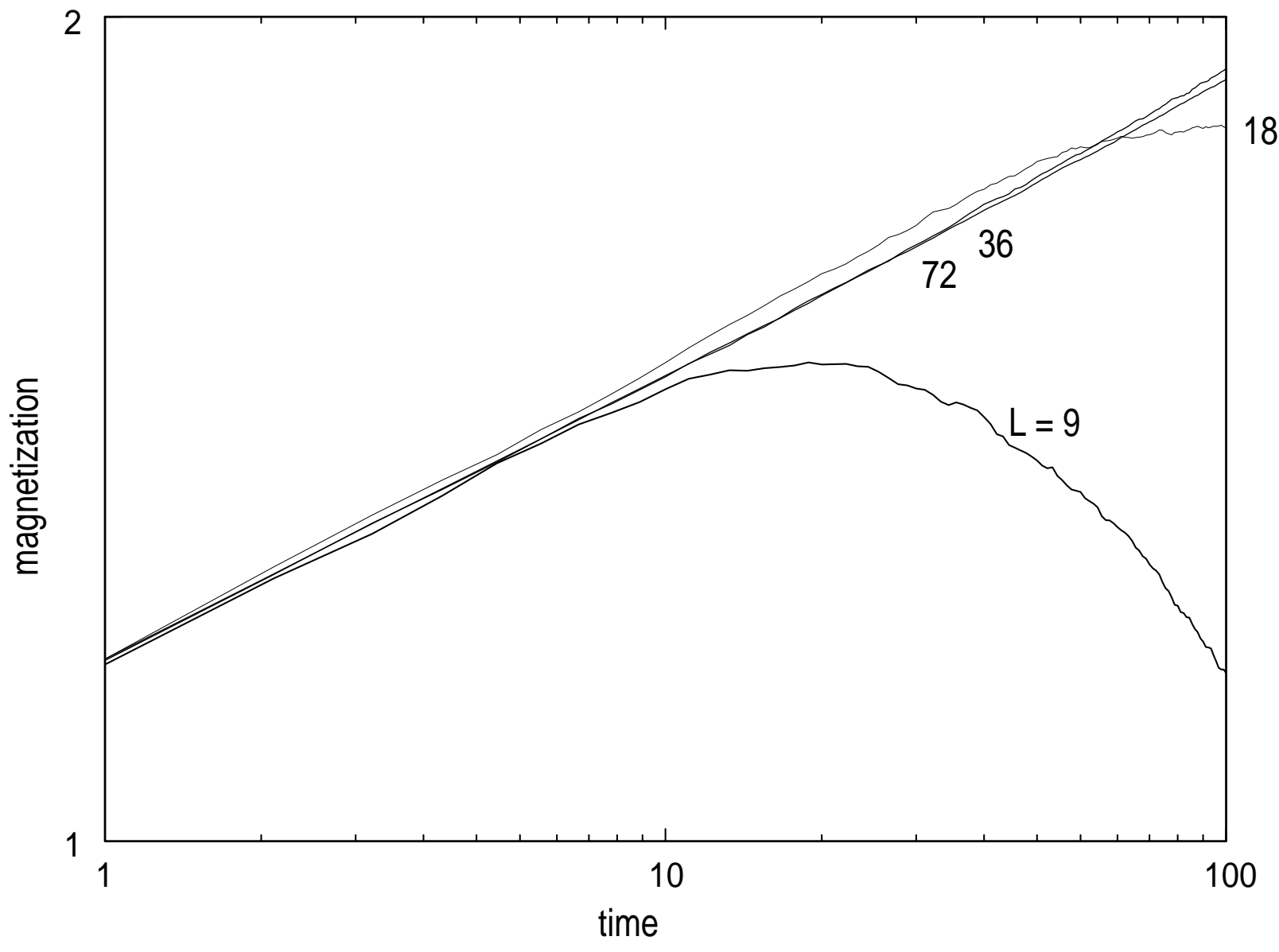
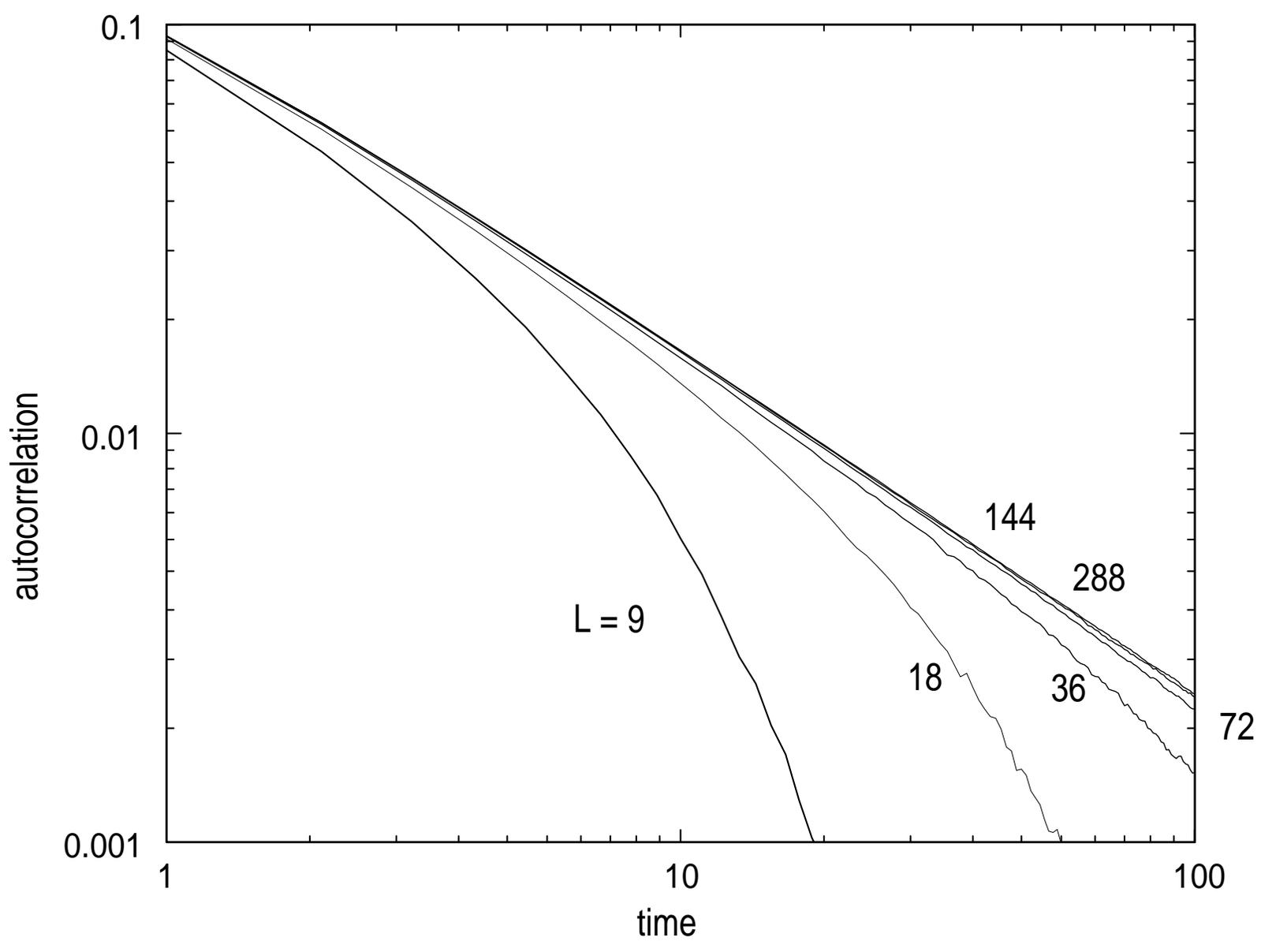
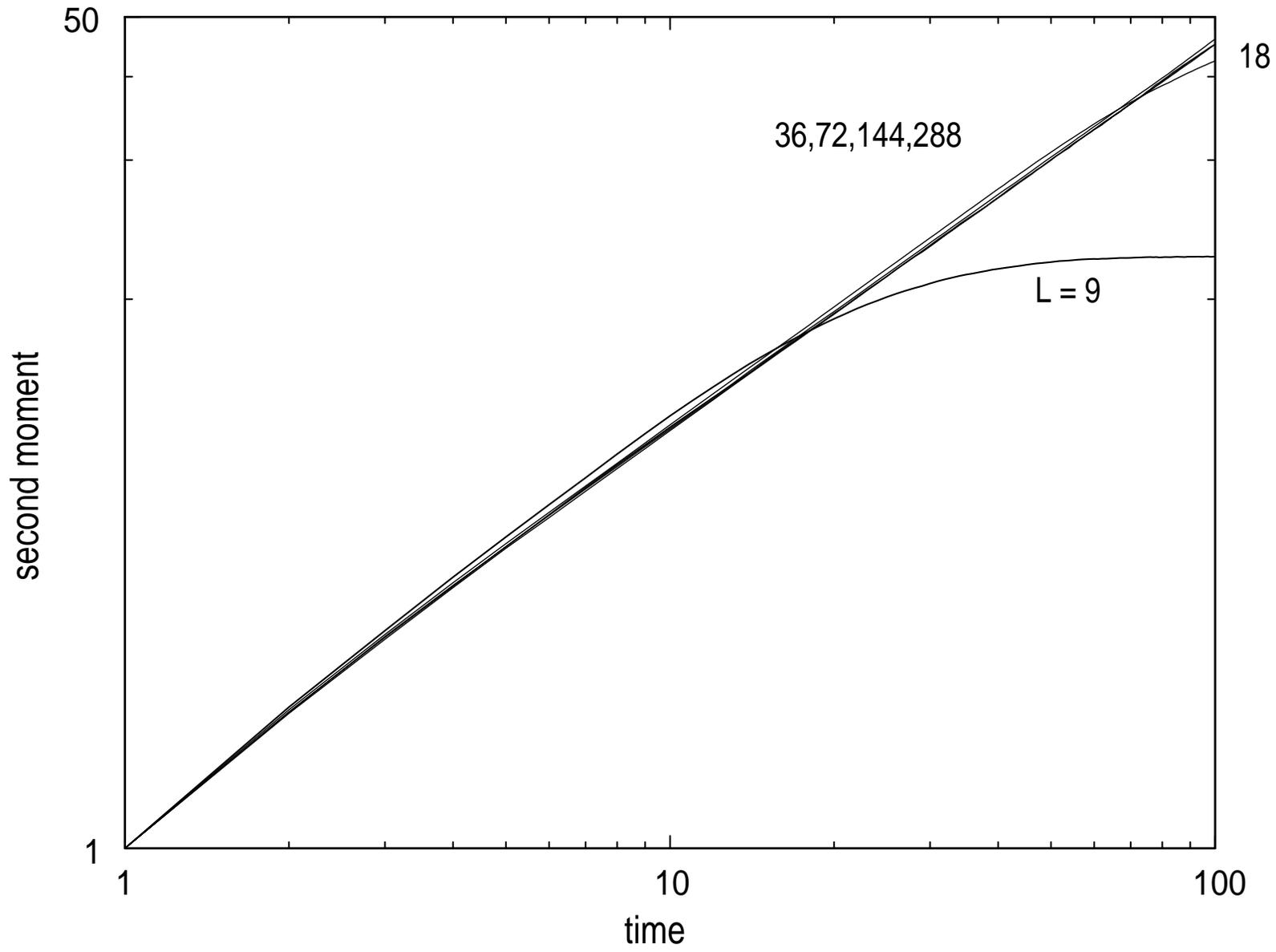
\end{document}